\documentclass[aps,prc,twocolumn,superscriptaddress]{revtex4-1}

\usepackage{url}
\usepackage{hyperref}
\usepackage{graphicx}
\usepackage{epstopdf}
\usepackage[outercaption]{sidecap}
\sidecaptionvpos{figure}{c}
\usepackage{xfrac}
\usepackage{amsmath}
\usepackage{amsfonts}

\begin{document}

\title{Improved precision on the experimental \emph{E}0 decay branching ratio of the Hoyle state}


\author{T.K.~Eriksen}
\altaffiliation[Current address: ]{Department of Physics, University of Oslo,\\N-0316 Oslo, Norway}
\affiliation{Department of Nuclear Physics, Research School of Physics,
    The Australian National University, Canberra, ACT 2601, Australia}
\author{T.~Kib\'{e}di}
\email{Corresponding author: Tibor.Kibedi@anu.edu.au}
\affiliation{Department of Nuclear Physics, Research School of Physics,
    The Australian National University, Canberra, ACT 2601, Australia}
\author{M.W.~Reed}
\affiliation{Department of Nuclear Physics, Research School of Physics,
    The Australian National University, Canberra, ACT 2601, Australia}
\author{A.E.~Stuchbery}
\affiliation{Department of Nuclear Physics, Research School of Physics,
    The Australian National University, Canberra, ACT 2601, Australia}
\author{K.J.~Cook}
\affiliation{Department of Nuclear Physics, Research School of Physics,
    The Australian National University, Canberra, ACT 2601, Australia}
\affiliation{Facility for Rare Isotope Beams, Michigan State University,
    640 South Shaw Lane, East Lansing, MI 48824, USA}
\author{A.~Akber}
\affiliation{Department of Nuclear Physics, Research School of Physics,
    The Australian National University, Canberra, ACT 2601, Australia}
\author{B.~Alshahrani}
\altaffiliation[Current address: ]{King Khalid University, Department of Physics,
    Faculty of Science, Abha 61413, Saudi Arabia}
\affiliation{Department of Nuclear Physics, Research School of Physics,
    The Australian National University, Canberra, ACT 2601, Australia}
\author{A.A.~Avaa}
\affiliation{iThemba LABS, National Research Foundation, P.O.~Box 722, 7129 Somerset West, South Africa}
\affiliation{School of Physics, University of Witwatersrand, Johannesburg, 2000, South Africa}
\author{K.~Banerjee}
\affiliation{Department of Nuclear Physics, Research School of Physics,
    The Australian National University, Canberra, ACT 2601, Australia}
\affiliation{Variable Energy Cyclotron Centre, 1/AF, Bidhan Nagar, Kolkata 700064 India}
\author{A.C.~Berriman}
\affiliation{Department of Nuclear Physics, Research School of Physics,
    The Australian National University, Canberra, ACT 2601, Australia}
\author{L.T.~Bezzina}
\affiliation{Department of Nuclear Physics, Research School of Physics,
    The Australian National University, Canberra, ACT 2601, Australia}
\author{L.~Bignell}
\affiliation{Department of Nuclear Physics, Research School of Physics,
    The Australian National University, Canberra, ACT 2601, Australia}
\author{J.~Buete}
\affiliation{Department of Nuclear Physics, Research School of Physics,
    The Australian National University, Canberra, ACT 2601, Australia}
\author{I.P.~Carter}
\affiliation{Department of Nuclear Physics, Research School of Physics,
    The Australian National University, Canberra, ACT 2601, Australia}
\author{B.J.~Coombes}
\affiliation{Department of Nuclear Physics, Research School of Physics,
    The Australian National University, Canberra, ACT 2601, Australia}
\author{J.T.H.~Dowie}
\affiliation{Department of Nuclear Physics, Research School of Physics,
    The Australian National University, Canberra, ACT 2601, Australia}
\author{M.~Dasgupta}
\affiliation{Department of Nuclear Physics, Research School of Physics,
    The Australian National University, Canberra, ACT 2601, Australia}
\author{L.J.~Evitts}
\altaffiliation[Current address: ]{Nuclear Futures Institute, Bangor University,
    Bangor, Gwynedd, LL57 2DG, United Kingdom}
\affiliation{TRIUMF, 4004 Wesbrook Mall, Vancouver, British Columbia V6T 2A3, Canada}
\affiliation{Department of Physics, University of Surrey, Guildford GU2 7XH, United Kingdom}
\author{A.B.~Garnsworthy}
\affiliation{TRIUMF, 4004 Wesbrook Mall, Vancouver, British Columbia V6T 2A3, Canada}
\author{M.S.M.~Gerathy}
\affiliation{Department of Nuclear Physics, Research School of Physics,
    The Australian National University, Canberra, ACT 2601, Australia}
\author{T.J.~Gray}
\affiliation{Department of Nuclear Physics, Research School of Physics,
    The Australian National University, Canberra, ACT 2601, Australia}
\author{D.J.~Hinde}
\affiliation{Department of Nuclear Physics, Research School of Physics,
    The Australian National University, Canberra, ACT 2601, Australia}
\author{T.H.~Hoang}
\affiliation{Research Center for Nuclear Physics, Osaka University, Ibaraki, Osaka, 567-0047, Japan}
\author{S.S.~Hota}
\affiliation{Department of Nuclear Physics, Research School of Physics,
    The Australian National University, Canberra, ACT 2601, Australia}
\author{E.~Ideguchi}
\affiliation{Research Center for Nuclear Physics, Osaka University, Ibaraki, Osaka, 567-0047, Japan}
\author{P.~Jones}
\affiliation{iThemba LABS, National Research Foundation, P.O.~Box 722, 7129 Somerset West, South Africa}
\author{G.J.~Lane}
\affiliation{Department of Nuclear Physics, Research School of Physics,
    The Australian National University, Canberra, ACT 2601, Australia}
\author{B.P.~McCormick}
\affiliation{Department of Nuclear Physics, Research School of Physics,
    The Australian National University, Canberra, ACT 2601, Australia}
\author{A.J.~Mitchell}
\affiliation{Department of Nuclear Physics, Research School of Physics,
    The Australian National University, Canberra, ACT 2601, Australia}
\author{N.~Palalani}
\altaffiliation[Current address: ]{University of Botswana, 4775 Notwane Rd., Gaborone, Botswana}
\affiliation{Department of Nuclear Physics, Research School of Physics,
    The Australian National University, Canberra, ACT 2601, Australia}
\author{T.~Palazzo}
\affiliation{Department of Nuclear Physics, Research School of Physics,
    The Australian National University, Canberra, ACT 2601, Australia}
\author{M.~Ripper}
\affiliation{Department of Nuclear Physics, Research School of Physics,
    The Australian National University, Canberra, ACT 2601, Australia}
\author{E.C.~Simpson}
\affiliation{Department of Nuclear Physics, Research School of Physics,
    The Australian National University, Canberra, ACT 2601, Australia}
\author{J.~Smallcombe}
\altaffiliation[Current address: ]{Oliver Lodge Laboratory, University of Liverpool,
    Liverpool L69 9ZE, United Kingdom}
\affiliation{TRIUMF, 4004 Wesbrook Mall, Vancouver, British Columbia V6T 2A3, Canada}
\author{B.M.A.~Swinton-Bland}
\affiliation{Department of Nuclear Physics, Research School of Physics,
    The Australian National University, Canberra, ACT 2601, Australia}
\author{T.~Tanaka}
\affiliation{Department of Nuclear Physics, Research School of Physics,
    The Australian National University, Canberra, ACT 2601, Australia}
\author{T.G.~Tornyi}
\altaffiliation[Current address: ]{Institute for Nuclear Research,
    The Hungarian Academy of Sciences, Debrecen 4026, Hungary}
\affiliation{Department of Nuclear Physics, Research School of Physics,
    The Australian National University, Canberra, ACT 2601, Australia}
\author{M.O.~de~Vries}
\affiliation{Department of Nuclear Physics, Research School of Physics,
    The Australian National University, Canberra, ACT 2601, Australia}

\date{\today}

\begin{abstract}
\begin{description}
\item[Background] Stellar carbon synthesis occurs exclusively via the $3\alpha$ process,
    in which three $\alpha$ particles fuse to form $^{12}$C in the excited Hoyle state,
    followed by electromagnetic decay to the ground state. The Hoyle state is above the
    $\alpha$ threshold, and the rate of stellar carbon production depends on the radiative
    width of this state.
    The radiative width cannot be measured directly, and must instead be deduced by combining
    three separately measured quantities.
    One of these quantities is the $E0$ decay branching ratio of the Hoyle state, and
    the current $10$\% uncertainty on the radiative width stems mainly from the uncertainty
    on this ratio.
    The rate of the $3\alpha$ process is an important input parameter in astrophysical
    calculations on stellar evolution, and a high precision is imperative to constrain
    the possible outcomes of astrophysical models.
\item[Purpose] To deduce a new, more precise value for the $E0$ decay branching ratio of the Hoyle state.
\item[Method] The $E0$ branching ratio was deduced from a series of pair conversion
    measurements of the $E0$ and $E2$ transitions depopulating the $0^+_2$ Hoyle state
    and $2^+_1$ state in $^{12}$C, respectively.
    The excited states were populated by the $^{12}$C$(p,p^\prime)$ reaction at 10.5~MeV
    beam energy, and the pairs were detected with the electron-positron pair spectrometer,
    Super-e, at the Australian National University.
    The deduced branching ratio required knowledge of the proton population of the two
    states, as well as the alignment of the $2^+_1$ state in the reaction.
    For this purpose, proton scattering and $\gamma$-ray angular distribution
    experiments were also performed.
\item[Results] An $E0$ branching ratio of $\Gamma^{E0}_{\pi}/\Gamma=8.2(5)\times10^{-6}$ was
    deduced in the current work, and an adopted value of
    $\Gamma^{E0}_{\pi}/\Gamma=7.6(4)\times10^{-6}$ is recommended based on a weighted
    average of previous literature values and the new result.
\item[Conclusions] The new recommended value for the $E0$ branching ratio is about
    14\% larger than the previous adopted value of $\Gamma^{E0}_{\pi}/\Gamma=6.7(6)\times10^{-6}$,
    while the uncertainty has been reduced from 9\% to 5\%.
    The new result reduces the radiative width, and hence $3\alpha$ reaction rate,
    by 11\% relative to the adopted value, and the uncertainty to 6.1\%.
    This reduction in width and increased precision is likely to constrain possible
    outcomes of astrophysical calculations.
\end{description}
\end{abstract}


\maketitle

\section{Introduction}
\label{sec:Intro}

 The synthesis of heavier elements in the universe is initiated by the $pp$-chain reactions
 in hydrogen burning stars, where four protons are ultimately converted into one $\alpha$
 particle with the release of energy.
 However, proton capture reactions forming heavier elements are inhibited by the rapid
 disintegration of $^{8}$Be, $\mathrm{T}_{1/2}=8.2\times 10^{-17}\mathrm{ s}$ \cite{2017Au03},
 into two $\alpha$ particles, so that no heavier elements are formed in stars at the hydrogen burning stage.
 It was not known how nucleosynthesis could proceed beyond $^{8}$Be until Salpeter
 suggested that an equilibrium concentration of $^{8}$Be can be sustained in a star of
 sufficient helium concentration and stellar temperature,
 resulting in a small probability for a third $\alpha$ particle to fuse with the $^{8}$Be and
 form $^{12}$C \cite{1952Salpeter_APJ}.
 Carbon production was thus suggested to occur via a sequential fusion of three $\alpha$
 particles, $\left(\alpha+\alpha\rightarrow \mathrm{^{8}Be}\right) +\alpha\rightarrow \mathrm{^{12}C}^*$,
 now commonly known as the $3\alpha$ process.
 The stellar conditions required for the $3\alpha$ process are fulfilled at the end of the
 hydrogen burning stage, due to gravitational contraction of the helium produced by the
 $pp$-chain reactions.
 The $0^+_2$ state at 7.65~MeV above the ground state in $^{12}$C is crucial for the
 $3\alpha$ process, as it acts as a resonance for $s$-wave $\alpha$ capture at the relevant
 stellar temperatures.
 Without this resonant state, the cross section for the sequential $3\alpha$ process would be
 too small to produce the observed carbon abundance in the universe.
 The resonant state was predicted by Fred Hoyle \cite{1953Ho81} before the first experimental
 observations \cite{1953Du23,1957Co59}, and became known as the Hoyle state.

\begin{figure}[h]
    \centering
    \includegraphics[width=\columnwidth]{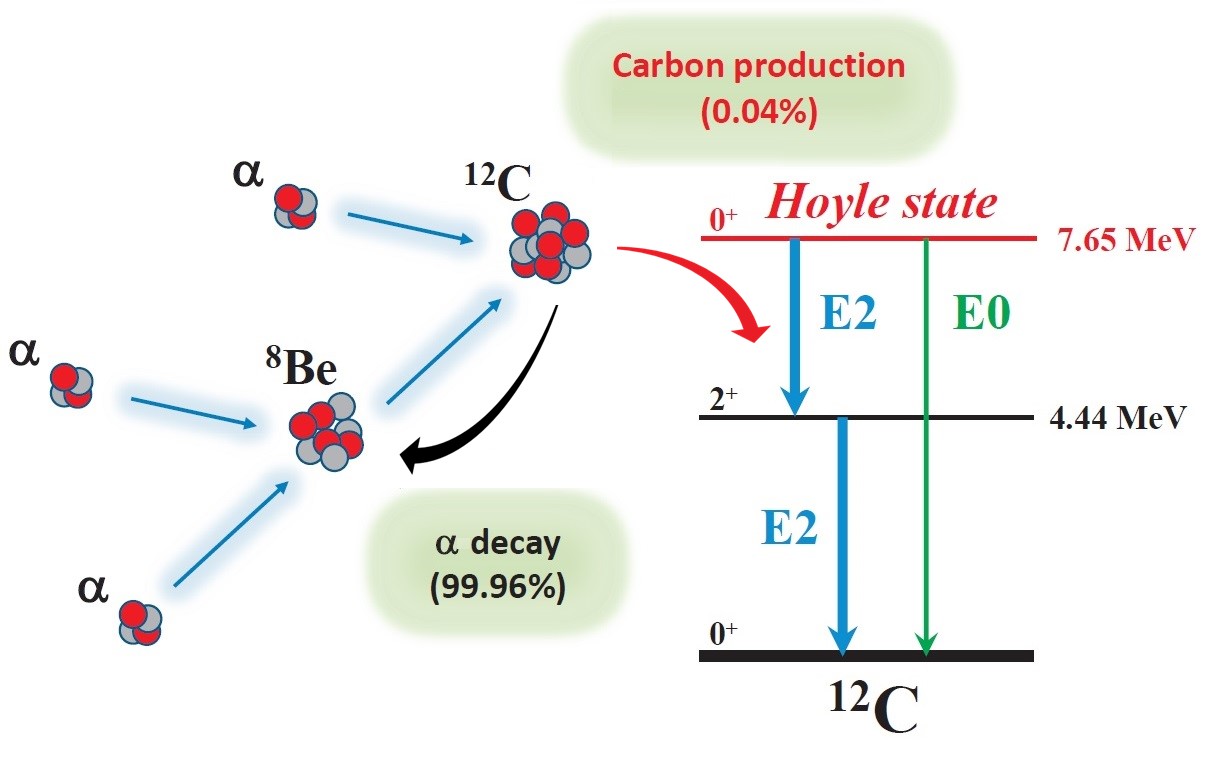}
    \caption{The $3\alpha$ process and the decay modes of the Hoyle state.}
    \label{fig:3aHoyle}
\end{figure}

 The Hoyle state energy exceeds the $\alpha$ decay threshold, and it disintegrates back to
 $^{8}\mathrm{Be}+\alpha$ or $3\alpha$ $\sim99.96\%$ of the time \cite{2014Fr09}.
 Stable carbon is only formed in $\sim0.04\%$ of the $3\alpha$ reaction instances, by
 electromagnetic decay to the ground state.
 Figure~\ref{fig:3aHoyle} provides a schematic illustration of the formation and various decay
 modes of the Hoyle state.
 Direct disintegration to three $\alpha$ particles occurs very rarely, as is indicated by recent
  measurements, which provide upper limits of 0.043\% \cite{2017De25,2017Sm03} and
  0.019\% \cite{2019Ra29} for this decay mode relative to the total $\alpha$ break-up.
  The branching ratio of direct vs.~sequential decay is important for structure studies
  of the Hoyle state, but it is not relevant in the context of stellar carbon formation
  because the contribution from direct fusion of three $\alpha$ particles is negligible.

 The carbon production rate can be described by the resonance equation \cite{1988RolfsRodney}
\begin{equation}
\label{eq:3aRr}
 r_{3\alpha} = 4\sqrt{27}\times \frac{N^{3}_{\alpha}\pi^3\hbar^5}{M^{3}_{\alpha}k^3_{\mathrm{B}}T^3}
 \times \frac{\Gamma_{\alpha} \Gamma_{\mathrm{rad}}}{\Gamma}\times e^{-(E_{3\alpha}/k_{\mathrm{B}}T)}~,
\end{equation}
 where $N_{\alpha}$ and $M_{\alpha}$ are the number density and mass of the interacting
 $\alpha$ particles, and $\Gamma$, $\Gamma_{\alpha}$, and $\Gamma_{\mathrm{rad}}$ are the
 total, $\alpha$ decay, and radiative decay widths of the Hoyle state, respectively.
 Furthermore, $E_{3\alpha}=0.38$~MeV is the energy released in the break-up of the Hoyle
 state, and $\hbar$, $k_{\mathrm{B}}$, and $T$ are the reduced Planck constant, the Boltzmann constant,
 and the temperature, respectively.
 Since the Hoyle state decays mainly by $\alpha$ emission, $\Gamma\approx\Gamma_\alpha$,
 and Eq.~(\ref{eq:3aRr}) may be simplified into the expression
\begin{equation}
\label{eq:3aRr_2}
 r_{3\alpha} \propto \frac{\Gamma_{\mathrm{rad}}}{T^3}\times e^{-(E_{3\alpha}/k_{\mathrm{B}}T)}~,
\end{equation}
 which shows that the carbon production rate depends directly on the radiative width of the Hoyle state.
 Due to the sequential nature of the $3\alpha$ process and the short half-life of $^{8}$Be,
 $\Gamma_{\mathrm{rad}}$ cannot be measured directly.
 However, it can be deduced indirectly by three independently measured quantities
 (shown in square brackets in Eq.~(\ref{eq:radwidth1})) according to
\begin{equation}
\label{eq:radwidth1}
 \Gamma_{\mathrm{rad}} = \left[\frac{\Gamma_{\mathrm{rad}}}{\Gamma}\right] \times
 \left[\frac{\Gamma}{\Gamma^{E0}_{\pi}}\right] \times \left[\Gamma^{E0}_{\pi}\right]~,
\end{equation}
 where $\Gamma^{E0}_{\pi}$ is the partial $E0$ pair decay width.
 The current recommended radiative width obtained from Eq.~(\ref{eq:radwidth1}) is
 $\Gamma_\mathrm{rad} = 3.7(4)$~meV \cite{2014Fr09}, which has an uncertainty of $10\%$.
 The uncertainties on the individual quantities are 2.5\%, 9.0\%, and 3.2\% for
 $\Gamma_{\mathrm{rad}}/{\Gamma}$
 \cite{1961Al23,1963Se23,1964Ha23,1974Ch03,1975Da08,1975Ma34,1976Ma46,1976Ob03},
 $\Gamma/\Gamma^{E0}_{\pi}$ \cite{1960Al04,1960Aj04,1972Ob01,1977Ro05,1977Al31}, and
 $\Gamma^{E0}_{\pi}$ \cite{2010Ch17},
 respectively, hence the uncertainty on the radiative width stems mainly from the challenges
 of measuring $\Gamma^{E0}_{\pi}/\Gamma$.
 The goal of the present work was to extract $\Gamma^{E0}_{\pi}/\Gamma$ by a new measurement
 with improved precision.

\section{Method}
\label{sec:Method}

 The $E0$ pair branching ratio of the Hoyle state was determined from electron-positron pair
 measurements of the ground state transitions of the first and second excited states in $^{12}$C,
 shown in Fig.~\ref{fig:3aHoyle}, based on the procedure reported by Alburger \cite{1977Al31}.
 In a $^{12}$C$(p,p^\prime)$ experiment, the number of experimentally measured $E0$ pairs following
 decay of the Hoyle state can be expressed as
\begin{equation}
    \label{eq:exppairintE0}
    N^{E0}_{\pi} = N_{p}(0^+_2) \times \frac{\Gamma^{E0}_{\pi}}{\Gamma} \times \epsilon^{E0}_{\pi}~,
\end{equation}
 where $N_{p}(0^+_2)$ is the number of protons populating the Hoyle state,
 $\Gamma^{E0}_{\pi}/\Gamma$ is the $E0$ pair decay branching ratio, and $\epsilon^{E0}_{\pi}$
 is the pair detection efficiency of the 7.65~MeV $E0$ transition.
 Similarly, for the $E2$ transition de-exciting the 4.44~MeV $2^+_1$ state in $^{12}$C,
 the expected number of pairs is
\begin{align}
    \label{eq:exppairintE2}
    N^{E2}_{\pi} =& \left[N_{p}(2^+_1) + \left( N_{p}(0^+_2) \times \frac{\Gamma_{\mathrm{rad}}(0^+_2\rightarrow 2^+_1)}{\Gamma}\right) \right] \\
    &\times \frac{\alpha_{\pi}}{(1+\alpha_{\pi})} \times \epsilon^{E2}_{\pi} \nonumber \\
    &\simeq N_{p}(2^+_1) \times \frac{\alpha_{\pi}}{(1+\alpha_{\pi})} \times \epsilon^{E2}_{\pi}~,\nonumber
\end{align}
 where the second term in the bracket may be omitted because
 $\Gamma \gg \Gamma_{\mathrm{rad}}(0^+_2\rightarrow 2^+_1)$.
 The pair decay probability of the 4.44~MeV $E2$ transition is accounted for by using the
 theoretical pair conversion coefficient, $\alpha_{\pi} = I_{\pi}/I_{\gamma}$.
 The $E0$ pair branching ratio of the Hoyle state may then be expressed by rearranging the ratio of
 Eq.~(\ref{eq:exppairintE0}) and Eq.~(\ref{eq:exppairintE2}) as
\begin{equation}
    \label{eq:G_Gpi1}
    \frac{\Gamma^{E0}_{\pi}}{\Gamma} = \frac{N^{E0}_{\pi}}{N^{E2}_{\pi}} \times \frac{N_{p}(2^+_1)}{N_{p}(0^+_2)} \times \frac{\epsilon^{E2}_{\pi}}{\epsilon^{E0}_{\pi}} \times \frac{\alpha_{\pi}}{(1+\alpha_{\pi})}~.
\end{equation}
 Hence, to deduce the $E0$ pair decay branching ratio one needs to measure the pair transitions
 and proton population of the two excited states in question.
 Furthermore, the angular distribution of the $E2$ $\gamma$-decay must be known to account
 for alignment of the $2^+_1$ state, which can affect the observed pair decay intensity.
 Measurements of the pair transitions and the proton population ratio of the two excited states,
 as well as of the angular distribution of the 4.44~MeV $\gamma$ ray de-exciting the $2^+_1$
 state, were performed in the present work.
 The detector efficiency for pair measurements was determined from Monte Carlo simulations,
 as described in Sec.~\ref{sec:SpecEff}.

\section{Experimental details}
\label{sec:Experiment}

\subsection{Spectrometer setup}
\label{sec:ExpSetup}
\begin{figure}[h]
    \centering
    \includegraphics[width=\columnwidth]{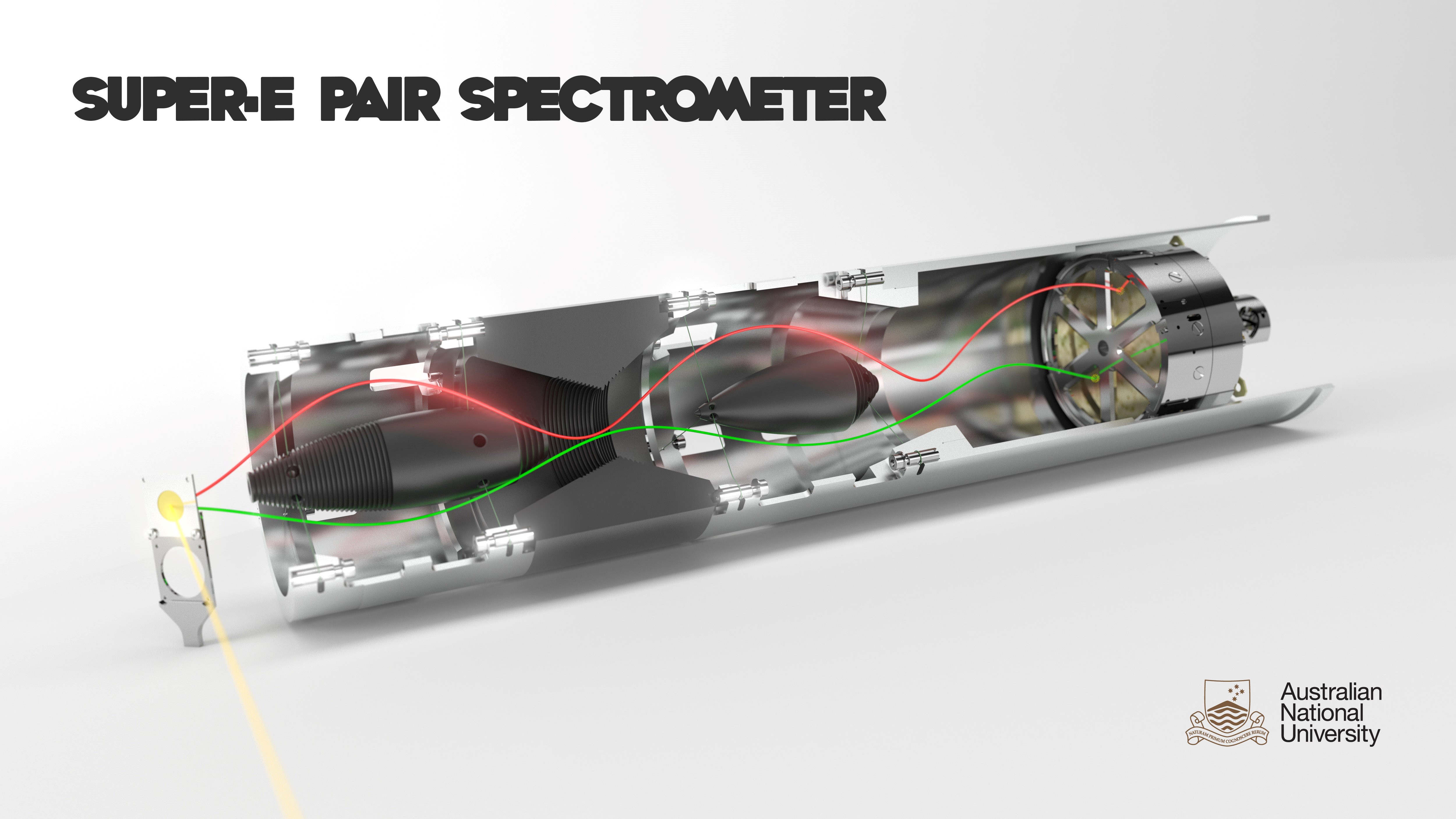}
    \caption{An illustration of the Super-e pair spectrometer, showing the target, baffle system, and detector array (from left to right). The setup is mounted perpendicular to the beam, which is represented in yellow. An electron-positron pair transmission is indicated by the red and green trajectories. Image courtesy of Thomas Tunningley, ANU.}
    \label{fig:Super-e}
\end{figure}

 The experimental setup is located in the Heavy Ion Accelerator Facility (HIAF) at The Australian National
 University (ANU).
 Proton beams were delivered by the 14~UD pelletron tandem accelerator \cite{HIAF}.
 A new spectrometer setup was developed and optimized for pair measurements, based on the existing ANU
 2.1~T superconducting solenoid \cite{1990Ki11}.
 The main upgrades involved a new baffle system and detector array, which will be described later
 in this section.
 The solenoid itself consists of liquid helium cooled NbTi coils, which provide a highly homogeneous
 and axially symmetric magnetic field, with a uniformity of
 $-3.7\%\leq\Delta B/B\leq+1.6\%$ within the spectrometer volume \cite{1990Ki11}.
 The coil current is computer controlled, and the magnetic field is monitored with a Hall probe.
 The most effective electron-positron pair measurements are achieved when the solenoid is set up to
 sample discrete magnetic fields providing maximum transmission of both pair constituents.
 The optimum magnetic fields depend on the transitions of interest and spectrometer transmission
 properties, as will be described in Sec.~\ref{sec:SpecEff}.
 The in-beam sampling is determined by integrated current in the beam dump.
 Figure~\ref{fig:Super-e} provides a cross-sectional illustration of the Super-e pair spectrometer,
 revealing its components.
 The spectrometer is mounted perpendicular to the beam axis, and its dimensions are defined by
 the solenoid bore diameter and the target-detector distance, which are \O~= 84.2~mm and
 $l=350$~mm, respectively.
 Starting from the left hand side of Fig.~\ref{fig:Super-e}, it can be seen that the target is
 positioned at 45$^\circ$ relative to the beam to allow electrons and positrons to be emitted
 through the rear of the target and into the spectrometer.
 Electrons and positrons emitted within the acceptance angles and momentum window of the spectrometer
 are transported through the baffle system and reach the detector plane after following helical
 trajectories due to the Lorentz force.
 The axially symmetric baffle system is designed to shield the detector array against $\gamma$ rays
 emitted from the target, and consists of two axial baffles and a diaphragm made of \emph{Heavymet}
 (W-Ni-Fe alloy) coated with a 1~mm layer of \emph{TorrSeal} (low vapor pressure epoxy),
 a low-$Z$ material intended to reduce both the amount of scattering and secondary electron production.

 The Si(Li) detector array, named Miel, consists of six identical, 9~mm thick sector-shaped Si(Li) segments,
 each with an active area of 236~mm$^2$ \cite{2012Kibedi_EPJ}.
 When assembled, the segments form an annular array, but are separated by 3-mm-thick, non-magnetic
 \emph{Heavymet} spacers to suppress cross-scattering of electrons and positrons between segments.
 Cross-scattering of  511~keV annihilation quanta is also suppressed.
 The assembled detector array can be seen to the right in  Fig.~\ref{fig:Super-e}.
 The Miel Si(Li) array may be operated as a single detector by summing the individual spectra of
 the segments, or in coincidence mode by requiring two or more segments to have fired, which is
 the case for the pair measurements in the present work.
 The six segments of Miel provide 15 unique two-segment coincidence combinations.
 The thickness of the segments allows for full absorption of electrons and positrons up to a
 kinetic energy of 3.5~MeV, which corresponds to a transition energy of 8~MeV for internal pair formation.
 Thus, the array is capable of detecting the 7.65~MeV $E0$ transition from the Hoyle state.

 The spectrometer setup is complemented by a HPGe detector used for monitoring the $\gamma$ emission
 from the target.
 The detector is positioned at 135$^\circ$ relative to the beam axis, 1.5 m away from the target,
 and has a crystal size of 81~mm $\times$ 54~mm (length $\times$ diameter).
 Data measured at different magnetic fields may then be normalized to relative sampling and
 reaction rates by using the peak area of a strong $\gamma$-ray transition in spectra projected
 with gates on the respective magnetic fields.
 The same $\gamma$ line is used for all normalizations in a particular experiment. In this work, the
 strong 4.44~MeV $2^+_1\rightarrow 0^+_1$ $\gamma$-ray transition was used for normalization.

 The quantities recorded in the current work were the energies and times from the six Si(Li) segments
 of Miel, the energy from the HPGe monitor detector, the solenoid control voltage and the
 Hall probe reading.
 There were two trigger requirements for storing the information, namely either two Si(Li) signals
 in coincidence or a signal from the HPGe monitor detector. The data were stored event-by-event,
 and sorted offline. Summed Miel energies, Miel time differences, and the magnetic rigidities of
 the particles were deduced from the stored quantities.
 The summed electron-positron pair energy could then be projected with gates on the physical
 momentum window of the spectrometer and prompt time differences, with background subtraction
 performed by gating on the random time differences.

\subsection{Spectrometer efficiency}
\label{sec:SpecEff}

\begin{figure}[h]
    \centering
    \includegraphics[width=0.939\columnwidth]{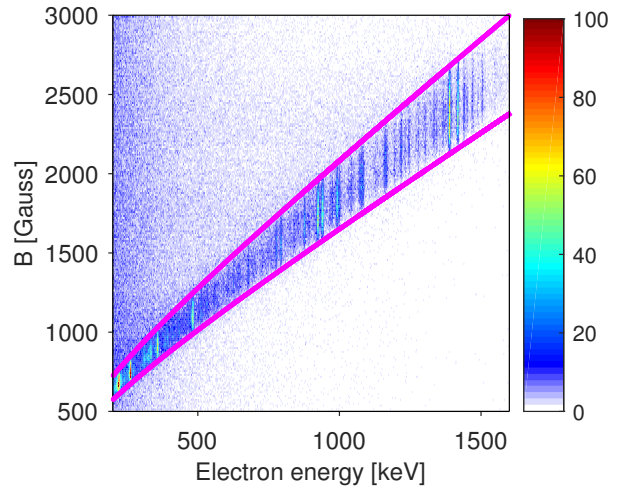}
    \caption{Energy vs.~magnetic field from a $^{170}$Lu source conversion-electron measurement,
    demonstrating the increasing width of the momentum window as a function of magnetic field
    strength and measured energy.
    The solid lines indicate the bounds of the momentum window, calculated as described in
    Ref.~\cite{1990Ki11}.
    The color scale indicates the number of counts.}
    \label{fig:EvsB}
\end{figure}

\begin{figure}[h]
    \centering
    \includegraphics[width=0.939\columnwidth]{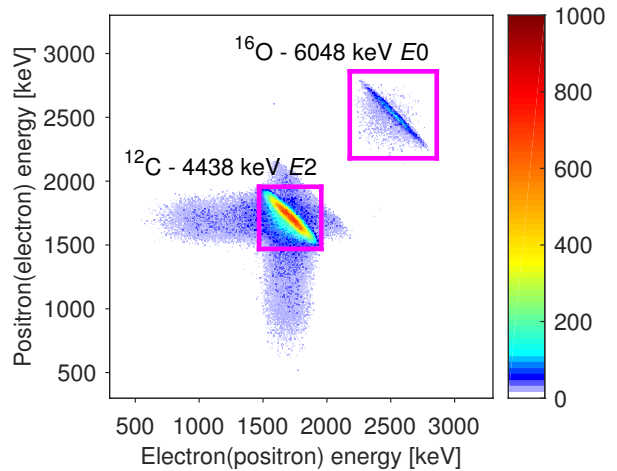}
    \caption{Energy vs.~energy from a $^{12}$C pair conversion measurement, showing the 4.44~MeV and
    6.05~MeV transitions in $^{12}$C and $^{16}$O, respectively.
    Note that the pair distribution for the 4.44~MeV transition is broadened due to the Doppler
    effect caused by decay from moving target recoils.
    The solid lines indicate the bounds of the momentum window, calculated as described in
    Ref.~\cite{1990Ki11}.
    The color scale indicates the number of counts.}
    \label{fig:EvsE}
\end{figure}

\begin{figure}[h]
    \centering
    \includegraphics[height=\columnwidth, angle=270]{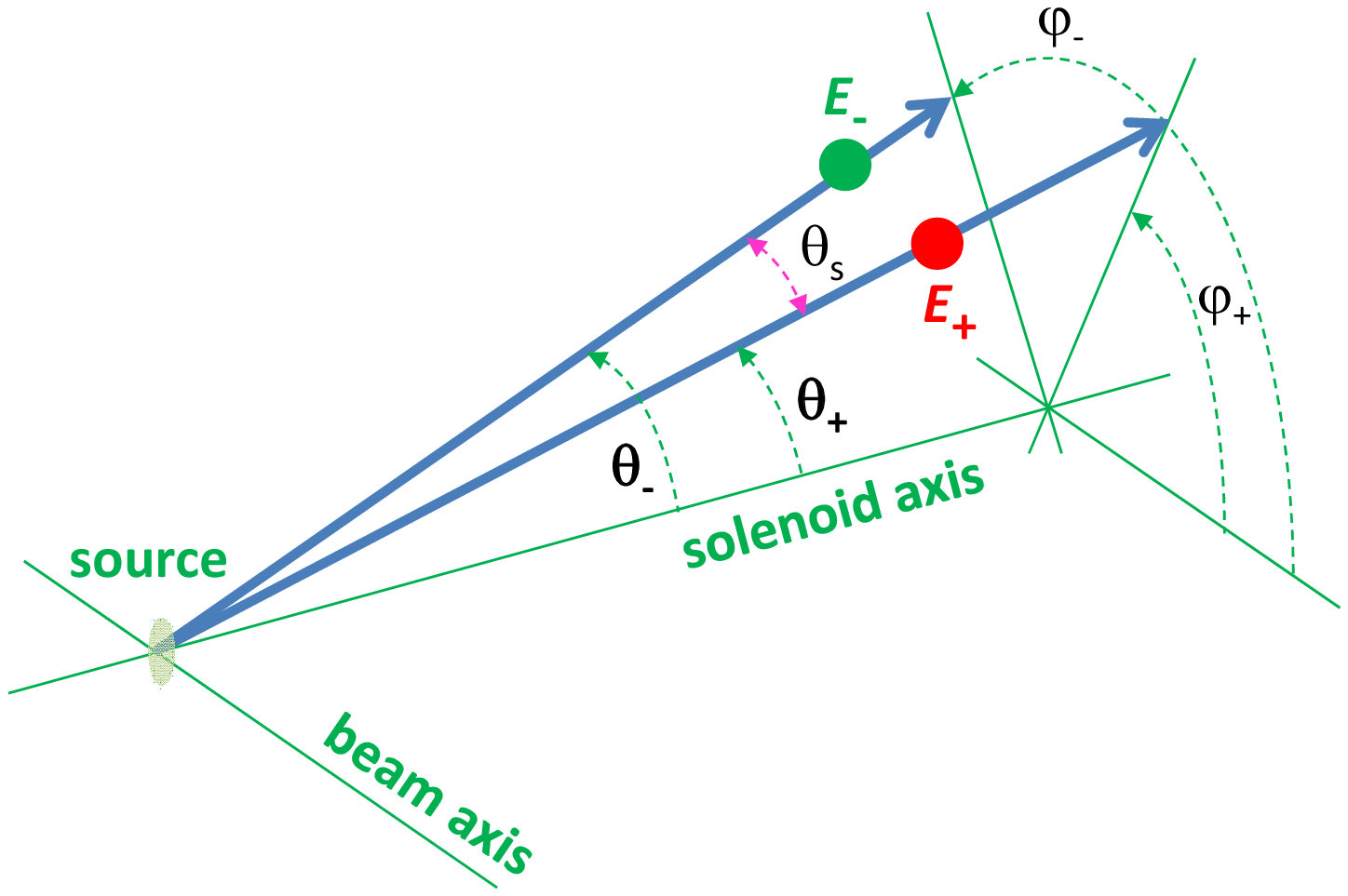}
    \caption{Pair emission in the spectrometer frame of reference.
    The intersection of the beam axis and spectrometer symmetry axis defines the origin of
    the coordinate system.}
    \label{fig:specframe}
\end{figure}

 The overall pair detection efficiency depends on the spectrometer transmission and intrinsic detector
  efficiency.
 The transmission is determined by the spectrometer acceptance angles with respect to the symmetry
 axis, $\theta\in[15.9^\circ, 46.9^\circ]$, the geometry of the baffle system, and the magnetic
 field strength.
 In addition to the directional limits of the acceptance angles, these properties define the
 physical limits in terms of momentum (the momentum window) for transportation of an electron or
 positron from the target through the baffle to the detector surface.
 Particles emitted within the acceptance angles and momentum window are able to reach the
 detector, while particles outside either will not be transmitted.
 The width and centroid of the momentum window increases with magnetic field strength,
 which means that the transmission efficiency of the spectrometer increases with particle energy,
 and that there is an optimum magnetic field for transportation of a certain particle energy.
 A magnetic field vs.~energy matrix from a singles conversion electron measurement is displayed
 in Fig.~\ref{fig:EvsB}, depicting the increasing momentum window as a function of magnetic
 field and measured energy.
 The solid lines indicate the limits of the momentum window.
 An example demonstrating the momentum window for pair measurements is provided in the energy vs.~energy
 matrix shown in Fig.~\ref{fig:EvsE}.
 The transmission of an electron-positron pair involves the directional kinematics of two correlated
  particles, for which the emission is dictated by the energy-angle correlation between the electron and positron.
 More specifically, the electron and positron share the available transition energy, less the
 energy consumed in the creation of two electron masses, $2m_0c^2$, according to the
 double-differential pair-emission probability.
 The double differential is defined as a function of positron energy, $E_+$, and separation
 angle of the pair, $\theta_s$, and depends on the transition energy and multipolarity.
 Figure~\ref{fig:specframe} illustrates the kinematics of a pair emission in the spectrometer
 frame of reference.
 In the present work, the double differential pair emission probability was calculated within
 the Born approximation with Coulomb correction, which will be explained in the following.
 Comparison of the distributions calculated with the Born approximation integrated over
 $\theta_s$, and single differential values for finite size calculations from
 Refs.~\cite{1979Sc31} and \cite{1990Ho21}, showed that the agreement was better
 than $1\%$ for $Z=6$ when $E_{-}\approx E_{+}$.
 Hence, the Born approximation was considered satisfactory for the $^{12}$C pair emission simulations.
 The double differential probability distribution for $E0$ pair transitions is determined by \cite{1941Oppenheimer}

\begin{align}
\label{eq:ddOE0}
 \frac{d^2\Omega_{\pi}(E0)}{dE_{+}d\mathrm{cos}\theta_s} =~& \\
 &p_{+}p_{-}\left( W_{+}W_{-} - m_0^2c^4 + p_{+}p_{-}c^2\mathrm{cos}\theta_s\right), \nonumber
\end{align}
 where $p_\pm$ denote momenta and $W_\pm=E_\pm+m_0c^2$ the total energies for electrons $(-)$ and
 positrons $(+)$.
 For higher electric multipoles, $EL$, the double differential distribution is given in terms of
 the pair conversion coefficient of the transition \cite{1949Ro19} by

\begin{widetext}
\begin{align}
\label{eq:ddaEL}
 &\frac{d^2\alpha_{\pi}(E L)}{dE_{+}d\mathrm{cos}\theta_s} = \left(\frac{2\alpha}{\pi(L+1)}\right) \left(\frac{p_{+}p_{-}}{q}\right) \frac{(q/\omega)^{2L-1}}{(\omega^2-q^2)^2}\times \Bigg[ (2L+1)\left( W_{+}W_{-}+1-\frac{p_{+}p_{-}}{3}\mathrm{cos}\theta_s\right) \\
 &~~~~~~+ L\left(\frac{q^2}{\omega^2}-2\right) (W_{+}W_{-}-1+p_{+}p_{-}\mathrm{cos}\theta_s)+ \frac{1}{3}(L-1)p_{+}p_{-} \left( \frac{3}{q^2}(p_{-}+p_{+}\mathrm{cos}\theta_s)(p_{+}+p_{-}\mathrm{cos}\theta_s)-\mathrm{cos}\theta_s\right) \Bigg]~, \nonumber
\end{align}
\end{widetext}
 where $\alpha$ is the fine structure constant, $q$ is the magnitude of the quantization vector,
 $\vec{q}=\vec{p_{+}}+\vec{p_{-}}$, and $\omega$ denotes the transition energy.
 It is important to note that in Eq.~(\ref{eq:ddaEL}), $\hbar=m_0=c=1$, so all energies are in
 terms of $m_0c^2$ and $p=\sqrt{W^2-1}$.

\begin{figure}[h]
    \centering
    \includegraphics[width=\columnwidth]{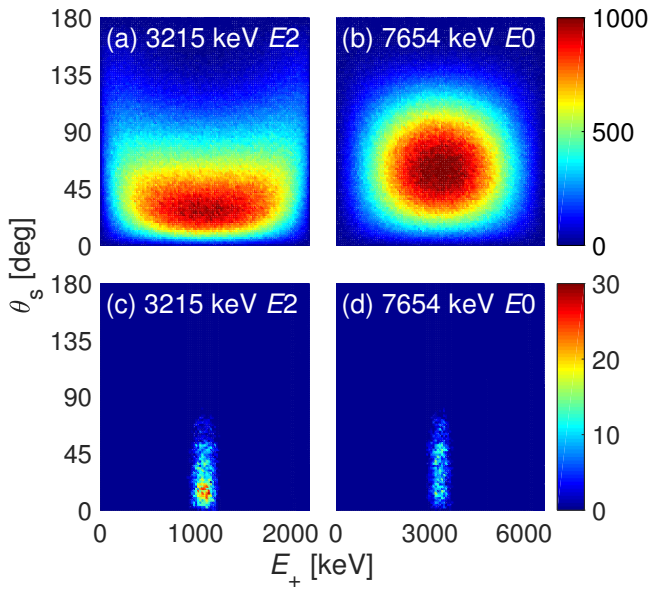}
    \caption{Monte Carlo simulations of pair emission and transmission. Panels (a) and (b): Double differential pair emission distributions for the transitions de-exciting the Hoyle state. The maximum emission probabilities are for $E_-\approx E_+$, and $\theta_s=30^\circ ~\mathrm{and}~ 60^\circ$ for the $E2$ and $E0$ transitions, respectively. \\ Panels (c) and (d): Detected pairs following emission according to the distributions in (a) and (b) and allowing for transmission through the spectrometer.}
    \label{fig:angcorr}
\end{figure}

 The evaluation of the pair transmission efficiency was performed using Monte Carlo simulations,
 by first simulating emission, and then transmission through the spectrometer.
 Pair emission was then sampled from the double differential probability distribution of
 Eq.~(\ref{eq:ddOE0}) for the $E0$ transition, and according to Eq.~(\ref{eq:ddaEL}) for the $E2$ transition.
 The distributions were corrected for Coulomb distortion of the emitted electron and positron energies,
 by multiplication with a correction factor as a function of positron energy.
 The Coulomb correction factor was estimated as described in Appendix~H in Ref.~\cite{1981Sc23}
\begin{equation}
\label{eq:ccf}
 F = \frac{(2\pi B_{+})(2\pi B_{-})}{(\mathrm{exp}(2\pi B_{+})-1)(1-\mathrm{exp}(-2\pi B_{-}))}~,
\end{equation}
 where $B_\pm$ denotes the relativistic Sommerfeld parameter $(Z\alpha E_\pm)/p_\pm$.
 The Coulomb correction, which also depends on the energy budget of the pair, is applied by
 multiplication with the distributions provided in Eqs.~(\ref{eq:ddOE0}) and (\ref{eq:ddaEL}).
 Double differential distributions calculated for the 3.22~MeV $E2$ and the 7.65~MeV $E0$ transitions
 from the Hoyle state in $^{12}$C are shown in Figs.~\ref{fig:angcorr} (a) and (b), respectively.
 Pairs emitted in the 4.44~MeV $E2$ transition are distributed in a similar fashion as shown in
 Fig.~\ref{fig:angcorr} (a), but with a different energy range. Since the 4.44~MeV $E2$ transition
 originates from the $2^+_1$ state of $^{12}$C, it is necessary to account for alignment of the
 nuclear spin states induced by the reaction and the effects on the corresponding pair emission
 distribution.
 The alignment correction is evaluated by using the distribution coefficients, $A_2$ and $A_4$, of
 the Legendre polynomials associated with the $\gamma$-ray angular distribution of the transition,
\begin{align}
\label{eq:gamma_angdist}
 W_\gamma(\theta_\mathrm{lab}) = A_0 + A_2P_2(\mathrm{cos}\theta_\mathrm{lab}) + A_4P_4(\mathrm{cos}\theta_\mathrm{lab})~, \nonumber \\
 &~
\end{align}
 where $P_\nu$ denotes the Legendre polynomial of order $\nu$, and $\theta_\mathrm{lab}$ is the
 $\gamma$-ray emission angle in the laboratory relative to the beam axis.
 The procedure for applying these coefficients to correct Eq.~(\ref{eq:ddaEL}) for alignment is
 explained in Refs.~\cite{1963Rose_PR,1964Wa24}.

 Trajectories of electrons and positrons emitted from the target were simulated by solving the
 relativistic equations of motion with the 4th order Runge-Kutta method.
 The equations were solved in a realistic magnetic-field profile for the solenoid calculated with
 Poisson Superfish \cite{Superfish}, and the particle trajectories were projected within a detailed
 specification of the spectrometer geometry in the spectrometer frame of reference.
 A trajectory calculation was terminated if the corresponding particle struck the surface of the
 absorber system or the inner bore.
 If both the electron and positron reached a Si(Li) segment, the event was registered as successful
 and all the parameters were stored.
 The pair-transmission efficiency was ultimately found by the ratio of pairs reaching two separate
 detector segments versus the number of emitted pairs.
 Transmitted and detected events of the emitted 3.22~MeV $E2$ and 7.65~MeV $E0$ pair transitions
 are shown in Figs.~\ref{fig:angcorr} (c) and (d), respectively.
 A potential cause of systematic uncertainty in the transmission efficiency would be from the use
 of the Born approximation with Coulomb correction, as opposed to applying calculations for
 extended nuclei.
 However, as mentioned above, the Born approximation differs by less than 1\% from the extended
 nuclei calculations performed for the low-$Z$ $^{12}$C nucleus, and no systematic uncertainties
 were assumed for the simulated transmission efficiency in the present work.

 The availability of sources for determining pair detection efficiencies, and even singles
 conversion electron detection efficiencies, is very limited.
 Consequently, the intrinsic detector efficiency was deduced from Monte Carlo simulations
 performed with the PENELOPE simulation tool \cite{PENELOPE}.
\begin{figure}[h]
    \centering
    \includegraphics[width=\columnwidth]{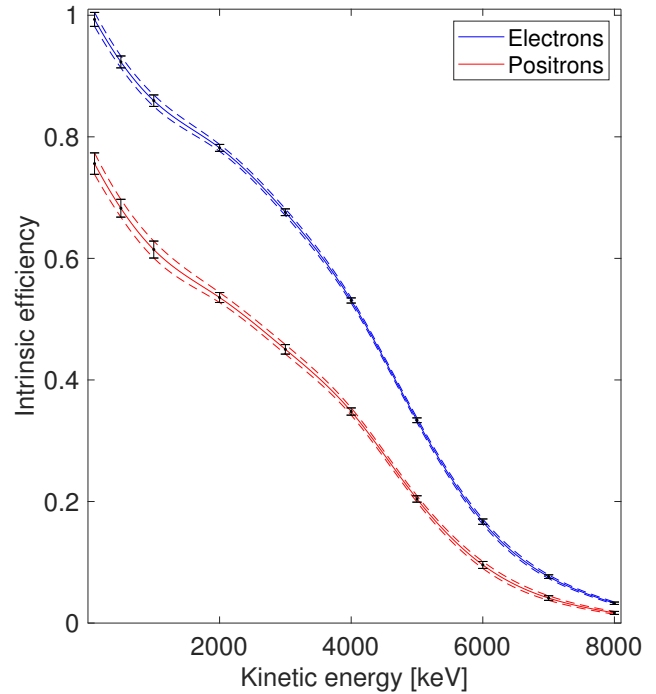}
    \caption{Intrinsic electron (blue/upper line) and positron (red/lower line) detection
     efficiencies of the Miel Si(Li) array, deduced from Monte Carlo simulations (data points).
     The error bars are defined as $3\sigma$ of the statistical uncertainties.
     The solid lines represent interpolations between data points, and the dashed lines
     indicate the uncertainties.}
    \label{fig:intrinsic}
\end{figure}
 Simulated spectra have previously been compared to $^{133}$Ba and $^{56}$Co conversion electron
 measurements \cite{2009Devlin}, with the conclusion that PENELOPE is reliable for the electron
 and positron energies relevant for the current work.
 Simulations of transmitted monoenergetic electrons and positrons between $0.1-8$~MeV were
 used as input for interaction simulations in conjunction with detailed geometry and material
 specifications of the detector array and spectrometer.
 The input parameters include realistic distributions of incident electron and positron angles,
 which are important for consideration of back-scattering.
 After folding in a resolution of 5~keV to the resulting response spectra, the intrinsic
 efficiency was deduced from the ratio of counts in the full energy peak versus the total
 number of counts in the spectrum.
 The intrinsic efficiencies obtained from the simulations are shown in Fig.~\ref{fig:intrinsic}.
 Note that the positron efficiency is in general lower than the corresponding electron efficiency.
 This is due to the fact that the positron-response spectra have an additional component above the
 full (kinetic) energy peak caused by energy deposition by annihilation radiation
 (converted from the rest mass).
 Hence, the ratio of counts in the full energy peak versus total counts is lower for positrons
 than it is for electrons.
 The interaction cross-sections of electrons and positrons in matter are otherwise essentially
 identical for the incident energies relevant to the pair measurements in the present work.

\subsection{Experimental conditions}
\label{sec:ExpConditions}

 The 4.44~MeV~2$^+_1$ and 7.65~MeV~0$^+_2$ levels in $^{12}$C were populated by using
 the $^{12}$C$(p,p^{\prime})$ reaction at 10.5~MeV proton energy, which is a resonant bombarding
 energy for population of the Hoyle state \cite{1971Da36}.
 Target foils of 1~mg/cm$^2$ and $2\times1$~mg/cm$^2$ natural carbon containing
 98.9\% $^{12}$C and 1.1\% $^{13}$C were used.
 The beam intensity varied between $0.5-1.0$~$\mu$A, but was mostly stable around $500-600$~nA.
 For the chosen target and beam energy, the cross sections for populating the 4.44~MeV and
 7.65~MeV levels are reported to be $\sigma_{4.44}=291$~mb \cite{1981Dy03} and
 $\sigma_{7.65}=86.5$~mb \cite{1971Da36}, respectively.
 The average energy losses of 10.5~MeV protons in the full thicknesses of the 1~mg/cm$^2$ and
 $2\times1$~mg/cm$^2$ target foils positioned at 45$^\circ$ relative to the beam are 56~keV
 and 110~keV \cite{SRIM}, respectively.
 A simple reaction rate calculation with 10.5~MeV monoenergetic protons and a beam intensity of
 500~nA impinging on a 1~mg/cm$^2$ target, yields rates of $r_{4.44}=6.94\times10^7$~s$^{-1}$
 and $r_{7.65}=2.06\times10^7$~s$^{-1}$ for population of the two excited states.
 By taking into account the relevant conversion coefficients, branching ratios, and spectrometer
 transmission, the rates of pair constituents striking different detector segments in coincidence
 were deduced.
 The deduced rates are $0.4$~pairs/min for the 3.22~MeV $E2$ transition, 4860~pairs/min for the
 4.44~MeV $E2$ transition and $5.5$~pairs/min for the 7.65~MeV $E0$ transition.
 These rates were calculated using optimum magnetic fields for transmission of pairs from the
 transitions, which are 0.20~T, 0.28~T, and 0.49~T, respectively.
 Furthermore, the target contained a small fraction of $^{16}$O, which allowed the 6.05~MeV $E0$
 pair transition from the 0$^+_2$ state to be sampled and conveniently used for energy calibration in conjunction with the strong 4.44~MeV transition in $^{12}$C. The optimum magnetic field for measuring the 6.05~MeV $E0$ transition was 0.40~T.
The magnetic field of the solenoid was stepped through the four discrete optimum field settings over several repeated cycles during each run. Each cycle had a duration of about 30 minutes, and the time spent at each magnetic field was controlled by the integrated beam current on the target. The amount of time allocated to each field was determined by the expected intensity of the transitions, with more time allocated to weaker transitions. In the present measurements, 68\%, 3\%, 3\%, and 26\% of a cycle was allocated to the 3.22~MeV, 4.44~MeV, 6.05~MeV, and 7.65~MeV transitions, respectively.

\section{Results}
\label{sec:Results}

\begin{figure}[t]
    \centering
    \includegraphics[width=\columnwidth]{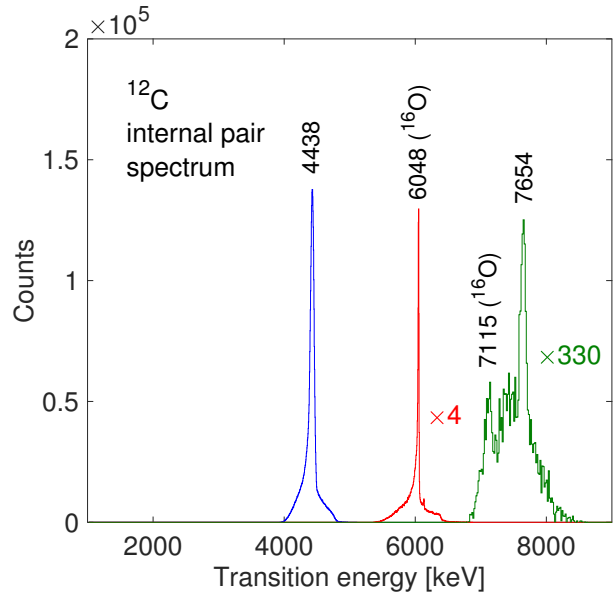}
    \caption{The summed pair spectrum of the three $^{12}$\textrm{C} experimental runs.
    The transitions are normalized to the peak area of the 4.44~MeV $\gamma$-ray transition
    measured by the monitor detector.
    Note that the 6048 keV and 7654 keV lines have been scaled up for visualization purposes.}
    \label{fig:pairspecs}
\end{figure}

 Four transitions were sampled during the pair measurements of the present work.
 They were the 4.44~MeV $2^+_1\rightarrow 0^+_1$, 3.22~MeV $0^+_2\rightarrow 2^+_1$, and
 7.65~MeV $0^+_2\rightarrow 0^+_1$ transitions in $^{12}$C, and the 6.05~MeV $0^+_2\rightarrow 0^+_1$
 transition in $^{16}$O.
 An initial objective was to detect the 3.22~MeV $E2$ pair transition from the Hoyle state, however,
 this turned out to be too ambitious as a large background rendered the observation of this
 weak transition impossible.
 Instead, the focus turned to the 4.44~MeV and the 7.65~MeV transitions, which were clearly visible.
 These two pair transitions in $^{12}$C, as well as the $^{16}$O line used for energy calibration,
 are shown in Fig.~\ref{fig:pairspecs}.
 Note that the spectrum has been shifted up in energy by $2m_0c^2=1022$~keV to reflect the transition energy.
 The spectrum in Fig.~\ref{fig:pairspecs} corresponds to 9 days of beam on target, from three experimental runs.
 To account for sampling time and beam intensity, the individual spectra were normalized to the peak area
 of the 4.44~MeV $\gamma$-ray transition measured by the monitor detector before summation.
 Furthermore, the spectra have been random subtracted by applying gates on prompt and random time differences.

\begin{figure}[h]
   \centering
    \includegraphics[width=\columnwidth]{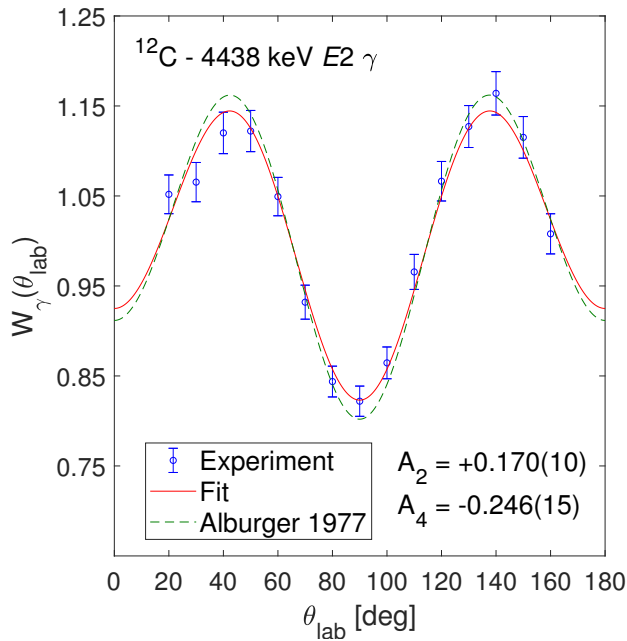}
    \caption{The angular distribution of $\gamma$ rays from the
    $2^+_1\rightarrow 0^+_1$ transition in $^{12}$C.
    The results are in good agreement with the previous measurement performed by Alburger \cite{1977Al31}.}
    \label{fig:Alignment4438keV}
\end{figure}

 Since the 4.44~MeV $E2$ transition originates from the $2^+_1$ state, the pair emission distribution
 for the transmission efficiency calculation had to be corrected for nuclear alignment effects.
 In order to obtain the distribution coefficients needed for the correction,
 the $\gamma$-ray intensities of the 4.44~MeV transition were measured at
 $\theta_\mathrm{lab} = 20^{\circ}-160^{\circ}$ in $10^{\circ}$ steps, using a HPGe detector with
 a crystal size of 81~mm $\times$ 54~mm (length $\times$ diameter) positioned 41.5~cm away
 from the target.
 The attenuation factors for this setup were found to be close to unity.
 For these measurements, a 1~mg/cm$^2$ thick natural carbon target was used, and the $2^+_1$
 state was populated by the $^{12}\mathrm{C}(p,p^{\prime})$ reaction at 10.5~MeV.
 The resulting angular distribution is shown with fitted distribution coefficients in
 Fig.~\ref{fig:Alignment4438keV}, and corresponds very well with the one measured
 by Alburger in 1977 \cite{1977Al31}.
 By comparing Monte Carlo simulations for pair transitions from unaligned and aligned cases of
 the 4.44~MeV state, a 7.45$\%$ reduction in transmission efficiency was revealed for the
 aligned case with the measured distribution coefficients.

 The proton population ratio, $N_{p}(2^+_1)/N_{p}(0^+_2)$, is also needed to extract
 $\Gamma^{E0}_{\pi}/\Gamma$ from the pair measurements according to the method described in
 Sec.~\ref{sec:Method}.
 For this reason, scattering measurements of $^{12}$C$(p,p^{\prime})$ were carried out using the
 ANU BALiN double sided silicon strip detector array \cite{2010Ra03,2011Luong_PLB,2016Cook_PhD}.
 The proton scattering distributions of the $2^+_1$ and $0^+_2$ states were measured simultaneously
 for scattering angles between $20^\circ-160^\circ$.
 Measurements were performed using both a 50~$\mu$g/cm$^2$ and the same 1~mg/cm$^2$ thick
 $^{12}$C target foil used in the pair conversion measurements.
 The 50~$\mu$g/cm$^2$ thick target was bombarded over several runs with proton beams of
 energies ranging between $10.4 - 10.7$~MeV, to obtain the angular distributions as a function
 of proton energy with little effect of energy loss in the target.
 The 1~mg/cm$^2$ thick target was bombarded with 10.5~MeV protons to obtain the proton
 angular distributions under the same conditions as in the $^{12}$C pair measurements of the
 present work.
 The angular distributions will be discussed in detail in a separate paper \cite{KCookAngDist}.
\begin{figure}[h]
    \centering
    \includegraphics[width=\columnwidth]{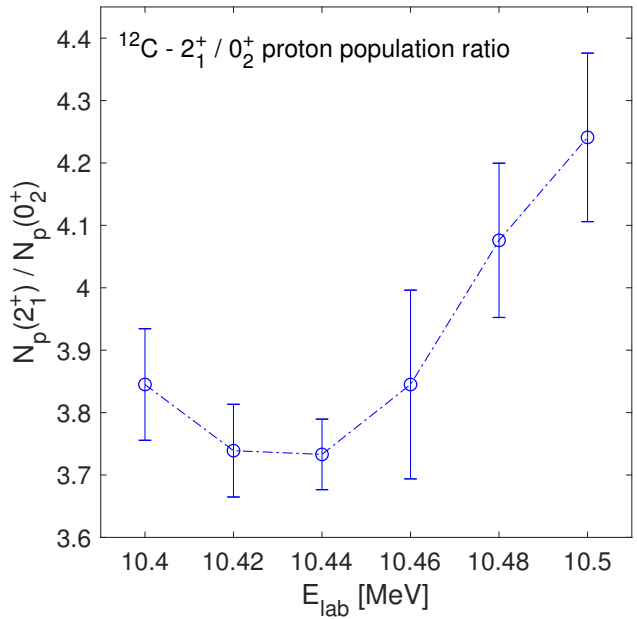}
     \caption{Proton population ratio of the $2^+_1$ and $0^+_2$ states in $^{12}$C as a
     function of proton beam energy.
     The dashed line represents a linear interpolation between the data points.}
    \label{fig:poprat_Edep}
\end{figure}
 Angular distribution functions were fitted to the data, and the ratio of the integrals over the
 full solid angle were used to deduce the proton population ratio of the $2^+_1$ and $0^+_2$ states.
  The 50~$\mu$g/cm$^2$ target measurements provided $N_{p}(2^+_1)/N_{p}(0^+_2)$ as a function
  of proton energy, which are shown in Fig.~\ref{fig:poprat_Edep} for energies relevant to
  the present work.
  By averaging the population ratio over the proton energy loss in the targets used in the pair
  measurements, proton population ratios of $N_{p}(2^+_1)/N_{p}(0^+_2)=3.97(12)$ and $3.89(10)$
  were obtained for the 1~mg/cm$^2$ and $2\times1$~mg/cm$^2$ thick targets, respectively.
 The corresponding ratio obtained from the proton scattering measurement with the 1~mg/cm$^2$ thick
 target yields $N_{p}(2^+_1)/N_{p}(0^+_2)=3.96(4)$, in excellent agreement with the averaged value.
 A weighted mean of $N_{p}(2^+_1)/N_{p}(0^+_2)=3.961(38)$ was adopted for the 1~mg/cm$^2$ target.
 These results are consistent with the previous value of $N_{p}(2^+_1)/N_{p}(0^+_2)=3.74(18)$
 from Alburger \cite{1977Al31}, which was deduced for a  3.5~mg/cm$^2$ thick target.

\begin{figure}[h]
    \centering
    \includegraphics[width=\columnwidth]{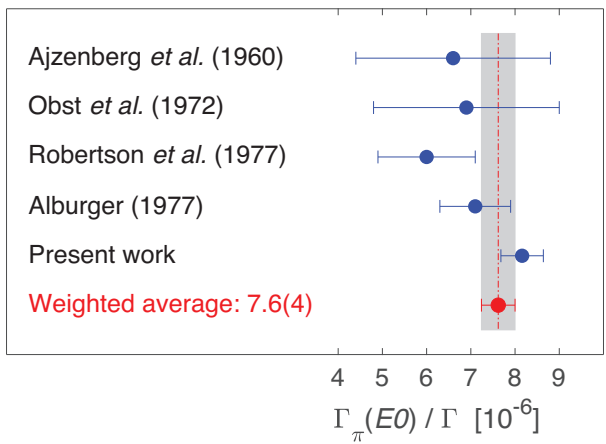}
    \caption{Previous, current, and weighted average values of $\Gamma^{E0}_\pi/\Gamma$.
    Further information about the previous measurements can be found in
    Refs.~\cite{1960Aj04,1972Ob01,1977Ro05,1977Al31} (listed in chronological order).}
    \label{fig:GE0_G}
\end{figure}

 The $E0$ pair branching ratio was deduced separately for the three pair measurements according to
 Eq.~(\ref{eq:G_Gpi1}),
 using the measured pair count ratios listed in the third column of Table~\ref{tab:12C_G0_G},
 and the relevant population ratio (2nd column), pair detection efficiencies, and the theoretical
 pair conversion coefficient \cite{2008Ki07,1979Sc31}.
 The efficiencies and conversion coefficient that
 were used in all three calculations are summarized in Table~\ref{tab:G0_G_calc}.
 The resulting $\Gamma^{E0}_{\pi}/\Gamma$ values are listed in the fourth column of
 Table~\ref{tab:12C_G0_G}.
 An average $E0$ pair decay branching ratio of $\Gamma^{E0}_{\pi}/\Gamma=8.2(5)\times10^{-6}$
 was found using \emph{AveTools} \cite{AveTools}, which utilizes three different methodologies
 to evaluate the average.
 These are the Limitation of Relative Statistical Weight, Normalized Residual Method, and the
 Rajeval Technique, which are explained in detail in Ref.~\cite{1992Ra08}.
 The three methods returned the same average value and uncertainty.
 A summary of the previous, current, and a weighted average of the $E0$ pair branching ratios is
 provided in Fig.~\ref{fig:GE0_G}.
 The weighted average in Fig.~\ref{fig:GE0_G} was also found using \emph{AveTools}.
 \begin{table}[h]
\caption{The experimental quantities used to deduce the $E0$ pair branching ratio of the Hoyle state,
and the resulting values.
The weighted average was obtained using \emph{AveTools} \cite{AveTools}.\label{tab:12C_G0_G}}
\begin{ruledtabular}
  \begin{tabular}{c c c c}
    Run & $\sfrac{N_{p}(2^+_1)}{N_{p}(0^+_2)}$  & $\sfrac{N^{E0}_\pi}{N^{E2}_\pi} \times 10^{4}$ & $\sfrac{\Gamma^{E0}_{\pi}}{\Gamma} \times 10^{6}$ \\
    \hline
    1 & 3.89(10)~ & 6.98(68)~ & 8.19(89)~ \\
    2 & 3.961(38) & 6.81(43)~ & 8.14(62)~ \\
    3 & 3.961(38) & 6.85(103) & 8.19(128) \\
    \hline
     ~ & ~ & Weighted average: & 8.2(5)~ \\
  \end{tabular}
\end{ruledtabular}
\end{table}
\begin{table}[h]
  \caption{The detection efficiencies and conversion coefficient used to deduce the $E0$ pair
  branching ratio of the Hoyle state.
  The pair conversion coefficient was obtained from BrIcc \cite{2008Ki07,1979Sc31}.\label{tab:G0_G_calc}}
\begin{ruledtabular}
  \begin{tabular}{c c c}
    $\epsilon^{E2}_{\pi}$ & $\epsilon^{E0}_{\pi}$ & $\alpha_{\pi}$ \\
    \hline
    $3.96(10)\times10^{-4}$ & $1.73(5)\times10^{-4}$ & $1.32(2)\times10^{-3}$ \\
  \end{tabular}
\end{ruledtabular}
\end{table}

\section{Discussion}
\label{sec:Discussion}

 As can be seen in Fig.~\ref{fig:GE0_G}, there are four previous published values for the
 $E0$ pair decay branching ratio of the Hoyle state.
 The results of Ajzenberg \emph{et al.}~\cite{1960Aj04} and Obst \emph{et al.}~\cite{1972Ob01}
 come from measurements of the neutron population ratio, $N_n(2^+_1)/N_n(0^+_2)$,
 in the reaction $^{9}$Be$(\alpha,n)^{12}$C at $E_\alpha=5.81$~MeV, while their results for the
 $E0$ pair branching ratio of the Hoyle state are both based on the pair intensity ratio,
 $N^{E0}_\pi/N^{E2}_\pi$, measured by Alburger in 1960 \cite{1960Al04} under the same
 experimental conditions.
 Robertson \emph{et al.}~\cite{1977Ro05} applied an independent and direct approach to
 deduce $\Gamma^{E0}_\pi/\Gamma$, by measuring the ratio of protons in coincidence with
 a 7.65~MeV pair transition over the singles proton rate
 $N^{E0}_{p,\pi}(0^+_2)/N^\mathrm{tot}_p(0^+_2)$ in a $^{12}$C$(p,p^{\prime})$
 experiment at $E_p=10.56$~MeV.
 The pair transitions were detected with a plastic scintillator detector covering nearly
 the full solid angle around the target, thus providing close to 100\% pair detection efficiency.
 However, due to the nature of the experimental setup, a number of corrections and uncertainties
 had to be considered in their analysis.
 In 1977, Alburger performed a pair intensity ratio measurement using the
 $^{12}$C$(p,p^{\prime})$ reaction at 10.5~MeV \cite{1977Al31}.
 The advantages of this approach are the resonant reaction for populating the Hoyle state,
  and the relative ease of measuring population ratios of protons as compared to neutrons.
  Alburger then deduced the $E0$ pair branching ratio according to the method described
  in Sec.~\ref{sec:Method}.
\begin{figure}[t]
    \centering
    \includegraphics[width=\columnwidth]{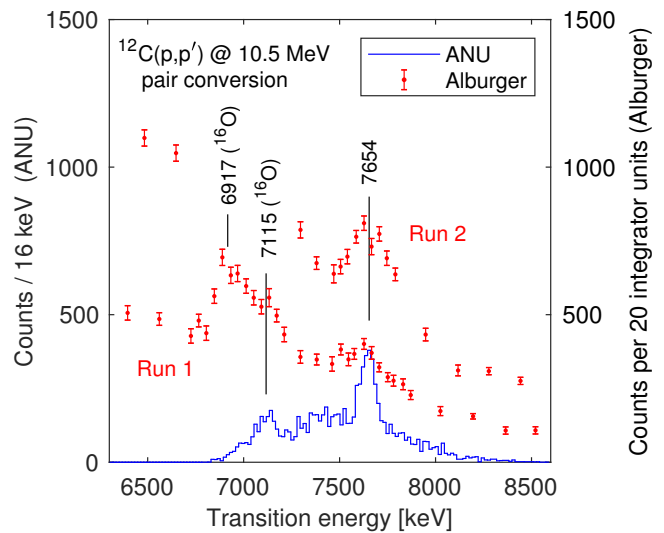}
    \caption{A comparison of the 7.65~MeV $E0$ pair spectra from the present measurement
    and Alburger's 1977 experiment \cite{1977Al31} to show the improved resolution and suppression of $^{16}$O peaks. The energy region containing the $^{16}$O peaks was excluded in Alburger's second run to save time.}
    \label{fig:compAl}
\end{figure}
 The same approach was adopted in the present work using data measured with the ANU Super-e spectrometer.
 The great improvement in resolving power of the present measurements, as compared to
 Alburger \cite{1977Al31}, is demonstrated in Fig.~\ref{fig:compAl}.
 The $E0$ pair branching ratio deduced from the present measurements,
 $\Gamma^{E0}_\pi/\Gamma = 8.2(5)\times10^{-6}$, agrees with that deduced by
 Alburger, $\Gamma^{E0}_\pi/\Gamma = 7.1(8)\times10^{-6}$, within the uncertainties.
 We recommend a weighted average of the previous and current measurements of the $E0$ pair
 decay branching ratio,  $\Gamma^{E0}_\pi/\Gamma = 7.6(4)\times10^{-6}$, for calculation
 of the radiative width of the Hoyle state.
 As a result, the present work reduces the uncertainty of the $E0$ pair branching ratio to
 5\% and increases its value by 14\% compared to the one adopted in the recent review by
 Freer and Fynbo \cite{2014Fr09}, $\Gamma^{E0}_\pi/\Gamma = 6.7(6)\times10^{-6}$.
 The new value of $\Gamma^{E0}_\pi/\Gamma$ provides a radiative width of
 $\Gamma_\mathrm{rad} = 3.28(20)$~meV when combined with the radiative branching ratio and
 $E0$ decay width reported in Ref.~\cite{2014Fr09},
 i.e.~$\Gamma_\mathrm{rad}/\Gamma=4.03(10)\times 10^{-4}$
 \cite{1961Al23,1963Se23,1964Ha23,1974Ch03,1975Da08,1975Ma34,1976Ma46,1976Ob03} and
 $\Gamma^{E0}_{\pi}=62.3(20)$~$\mu$eV \cite{2010Ch17}.
 Compared to the previously adopted value of the radiative width of the Hoyle state,
 $\Gamma_\mathrm{rad} = 3.7(4)$~meV \cite{2014Fr09}, the new result for the width
 agrees within the error bars, but is 11\% smaller and has an uncertainty of 6.1\% as compared to 10\%.

\begin{figure}[t]
    \centering
    \includegraphics[width=\columnwidth]{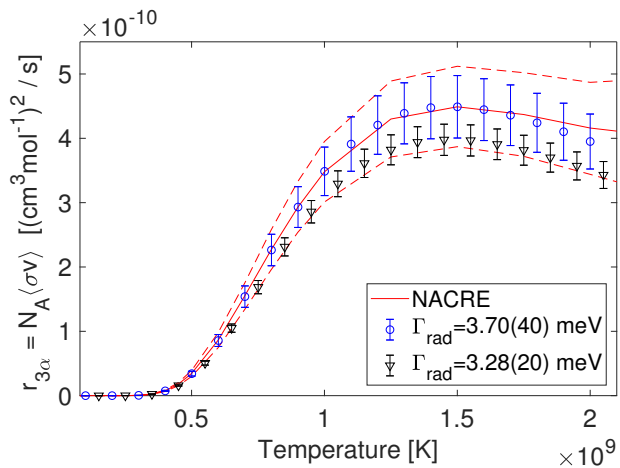}
    \caption{The $3\alpha$ reaction rate calculated within the temperature range of helium
    burning red giant stars using the NACRE library value \cite{1999An35}
    (solid red line with dashed lines indicating the range of uncertainty), previous
    recommended value (blue circles), and new recommended value (black triangles) of the radiative
    width of the Hoyle state.}
    \label{fig:r3a}
\end{figure}

 Plots of the estimated $3\alpha$ reaction rates, $r_{3\alpha}$, using the previous and
 current radiative widths are provided in Fig.~\ref{fig:r3a}.
 The figure also includes rates calculated with the standard NACRE library value \cite{1999An35}.
 The reaction rates agree well within the uncertainties, and it is clear that the new value
 on $\Gamma_\mathrm{rad}$ would not significantly change our astrophysical models and predictions.
 However, the reduced uncertainty will constrain possible scenarios and outcomes of the calculations,
 and facilitates advances in the research on stellar evolution and element synthesis in the universe.
 A major implication of $r_{3\alpha}$ is its effect on the carbon-to-oxygen abundance ratio at
 the end of the helium burning phase of stars, in which the
 $3\alpha$ process and $^{12}\mathrm{C}(\alpha,\gamma) ^{16}\mathrm{O}$ reaction compete
 for the available $\alpha$ particles, with the latter reaction also feeding on the
 available $^{12}$C nuclei.
 The carbon-to-oxygen abundance ratio is important for later stages of stellar evolution, and
 the rates of production and consumption of $^{12}$C are therefore important input parameters
 in astrophysical calculations.

 A recent measurement \cite{2020Kibedi_PRL} of the radiative branching ratio, $\Gamma_\mathrm{rad}/\Gamma$,
  suggests a value that is substantially higher than the currently adopted ratio used in this work.
  Combining this recent result with the present measurement on $\Gamma^{E0}_\pi/\Gamma$, results
  in a large increase of the radiative width as compared to the adopted value.
  This increase would have a significant impact on astrophysical calculations, and it is crucial to
  address the discrepancy observed for $\Gamma_\mathrm{rad}/\Gamma$.

 A new approach to determine the radiative width from a direct measurement of the ratio of the pair
 transitions de-exciting the Hoyle state, $\Gamma^{E2}_\pi/\Gamma^{E0}_\pi$, has been developed \cite{2012Kibedi_EPJ}.
 However, the success of this new method requires a 20 times reduction in the background currently
 observed in vicinity of the 3.22~MeV $E2$ pair peak.
 If this can be done, this approach has the potential to provide an independent measurement
 and settle the discrepancy for $\Gamma_\mathrm{rad}$. \newline\vfill

\begin{acknowledgments}
 The project was supported by the Australian Research Council Discovery Grants
 DP140102986, DP170101673, and DP170102423.
 Operation of the ANU Heavy Ion Accelerator Facility is supported by the NCRIS HIA capability.
 The support from technical staff for the development of the pair spectrometer, as well as
 during the long experimental runs, is greatly appreciated.\\
\indent This work was partially supported by:
 \\The International Joint Research Promotion Program of Osaka University and JSPS KAKENHI
    Grant Number JP 17H02893.
 \\The Natural Sciences and Engineering Research Council of Canada.
 \\The National Research Foundation (NRF), South Africa, under grant numbers [93533 and 118645].
\end{acknowledgments}



\end{document}